\documentclass[prd,amsmath,amssymb,nofootinbib,superscriptaddress,twocolumn,10pt]{revtex4}%superscriptaddress,showpacs,showkeys,

\usepackage{graphicx}
\usepackage{dcolumn}
\usepackage{bm}
\usepackage{amssymb}
\usepackage{latexsym}
\usepackage{booktabs}
\usepackage{amsmath}
\usepackage{multirow}
\usepackage{enumerate}
\usepackage{url}
\usepackage{subfigure}
\usepackage{float}
\usepackage[colorlinks=true, linkcolor=red, citecolor=blue]{hyperref}

\begin{document}

% \title{Multiband observations of intermediate-mass binary black holes with Taiji: detection prospects and cosmological forecasts}
\title{Cosmological prospects for multiband detection of intermediate-mass binary black holes with Taiji and ground-based detectors}

\author{Yue-Yan Dong}
\affiliation{Liaoning Key Laboratory of Cosmology and Astrophysics,
College of Sciences, Northeastern University, Shenyang 110819, China}
\author{Ji-Yu Song}
\affiliation{Liaoning Key Laboratory of Cosmology and Astrophysics,
College of Sciences, Northeastern University, Shenyang 110819, China}
\author{Jing-Fei Zhang}\thanks{Corresponding author; jfzhang@mail.neu.edu.cn}
\affiliation{Liaoning Key Laboratory of Cosmology and Astrophysics,
College of Sciences, Northeastern University, Shenyang 110819, China}
\author{Xin Zhang}\thanks{Corresponding author; zhangxin@mail.neu.edu.cn}
\affiliation{Liaoning Key Laboratory of Cosmology and Astrophysics,
College of Sciences, Northeastern University, Shenyang 110819, China}
\affiliation{MOE Key Laboratory of Data Analytics and Optimization for Smart Industry,
Northeastern University, Shenyang 110819, China}
\affiliation{National Frontiers Science Center for Industrial Intelligence and Systems Optimization,
Northeastern University, Shenyang 110819, China}

\begin{abstract}

Intermediate-mass black holes (IMBHs) bridge the gap between stellar-mass and supermassive black holes, but remain challenging to detect electromagnetically. Gravitational-wave observations provide a direct means of detecting IMBHs and their mergers. We simulate the gravitational-wave signals of IMBH binaries under different population models and assess their detectability with the space-based detector Taiji alone and in a multiband network combining Taiji with third-generation ground-based detectors. Taiji performs well in detecting high-mass IMBH binaries, while ground-based detectors compensate for its reduced sensitivity to lower-mass systems. Their combination expands the accessible parameter space and improves the constraints on cosmological parameters. In particular, multiband observations improve the constraint accuracy on $H_0$ by $36.5\%$ and $31.0\%$ compared with Taiji and ET2CE alone, respectively. We further examine the dependence of parameter accuracy on the number of simulated events, finding that improvements are most pronounced for small samples and gradually saturate as the number of events increases. We conclude that multiband observations enhance the detectability of IMBH binaries and reinforce their role as probes of precision cosmology.

\end{abstract}

\maketitle

\section{Introduction}\label{sec:intro}

In 2015, the LIGO Scientific Collaboration reported the first detection of gravitational waves (GWs) from a binary black hole merger, GW150914 \cite{LIGOScientific:2016dsl}, marking the beginning of GW astronomy and demonstrating that GW observations provide an independent probe of the Universe.
During the first three observing runs \cite{LIGOScientific:2018mvr,LIGOScientific:2020ibl,LIGOScientific:2021djp}, the detected events were dominated by mergers of stellar-mass binary black holes. In the latest fourth observing run \cite{LIGOScientific:2025slb}, however, the event GW231123 \cite{LIGOScientific:2025rsn} involved the merger of black holes with component masses of approximately $100$--$140\,M_\odot$, yielding a remnant black hole of about $225\,M_\odot$ and thereby hinting at the possible existence of intermediate-mass black holes (IMBHs).

IMBHs with masses typically in the range $10^{2}$--$10^{5}\,M_\odot$ constitute a transitional population bridging stellar-mass and supermassive black holes, and are expected to play an important role in the formation and evolution of galaxies, potentially serving as seeds of supermassive black holes \cite{Greene:2019vlv, Madau:2001sc, Silk:2017yai, Natarajan:2020avl, Chen:2025uzf}. However, detecting IMBH binaries with ground-based GW detectors remains challenging, since these detectors are primarily sensitive to frequencies in the tens of hertz to kilohertz band, whereas the GW signal of IMBH binaries typically lies below this band, thereby limiting observational constraints on their formation and evolution.

In the 2030s, space-based GW detectors are expected to operate in the millihertz band, thereby opening a low-frequency window that is largely inaccessible to ground-based instruments \cite{LISA:2017pwj, Robson:2018ifk, LISACosmologyWorkingGroup:2022jok, TianQin:2015yph, Liu:2020eko, Luo:2020bls, Milyukov:2020kyg, TianQin:2020hid, Hu:2017mde}. Their sensitivity to low-frequency GW signals enables long-duration observations of intermediate-mass binary black holes (IMBBHs) well before merger, providing information on their early orbital evolution. By combining space-based and ground-based detectors in a multiband observation, the signal evolution can be tracked over a broader frequency range, significantly improving sky localization and parameter estimation, which are crucial for astrophysical interpretation of IMBHs and for their application as dark sirens in cosmology.

GW sources can serve as standard sirens because their waveforms allow a direct inference of the luminosity distance \cite{Schutz:1986gp,Holz:2005df}. In light of the more-than-$5\sigma$ discrepancy, known as the Hubble tension \cite{Verde:2019ivm,Riess:2019qba,DiValentino:2021izs,Perivolaropoulos:2021jda,Abdalla:2022yfr,Kamionkowski:2022pkx}, between model-independent distance ladder measurements \cite{Breuval:2024lsv, Riess:2021jrx} and the Planck 2018 CMB inference of $H_0$ under the $\Lambda$CDM model \cite{Planck:2018vyg}, independent $H_0$ measurements based on GW standard sirens are particularly valuable \cite{Dalal:2006qt, Cutler:2009qv, Nissanke:2009kt, Zhao:2010sz, Cai:2016sby, Cai:2017aea, Du:2018tia, Belgacem:2019tbw, Safarzadeh:2019pis, Yang:2019bpr, Bachega:2019fki, Li:2024qso, Wang:2018lun, He:2019dhl, Chen:2020dyt, Hogg:2020ktc, Chen:2020zoq, Mitra:2020vzq, Qi:2021iic, Bian:2021ini, Ye:2021klk, deSouza:2021xtg, Califano:2022syd, Yang:2023zxk, Song:2025ddm, Song:2025bio, Jin:2025dvf, Wang:2021srv, Zhang:2019ylr, Li:2023gtu, Du:2025odq}. To date, the latest O4 result, based on GW170817 \cite{LIGOScientific:2017vwq} and a number of dark siren events, yields \(H_0 = 76.6^{+13.0}_{-9.5}\ \mathrm{km\,s^{-1}\,Mpc^{-1}}\) \cite{LIGOScientific:2025jau}. However, the constraint remains relatively weak due to limited distance precision and the scarcity of electromagnetic counterparts. For dark sirens, the lack of counterparts makes accurate sky localization and distance estimation especially important. Multiband observations can significantly improve localization and distance measurement accuracy \cite{Sesana:2016ljz,Vitale:2016rfr,Sesana:2017vsj,Isoyama:2018rjb,Jani:2019ffg,Carson:2019kkh,Liu:2020nwz,Datta:2020vcj,Zhang:2021pwe,Nakano:2021bbw,Yang:2021qge,Muttoni:2021veo,Liu:2021dcr,Zhu:2021bpp,Kang:2022nmz,Klein:2022rbf,Seymour:2022teq,Baker:2022eiz,Zhao:2023ilw, Dong:2024bvw, Song:2026kii}, which motivates the exploration of new source populations and observing strategies. In this regard, IMBBHs are particularly promising, as they are detectable in both space-based and ground-based frequency bands and are thus naturally suited for multiband cosmology.

Taiji is a space-based GW mission \cite{Hu:2017mde} that provides an important platform for multiband observations. As a space-based detector with high sensitivity at low frequencies, extensive efforts have been devoted in recent years to developing its scientific objectives and assessing its potential across a broad range of GW sources \cite{Ruan:2018tsw, Luo:2019zal, Ruan:2020smc, Luo:2021qji, Orlando:2020oko, Wang:2021srv, Zhao:2019gyk, TaijiScientific:2021qgx, Jin:2021pcv, Jin:2023sfc, Ren:2023yec}.
In this work, we consider a joint observational network consisting of the Taiji space-based detector and third-generation (3G) ground-based GW detectors. We systematically analyze the performance of multiband observations for IMBBH mergers, and assess the prospects of using these systems as dark sirens to constrain cosmological parameters.
We first assess the detectability of binary black hole systems with Taiji alone and with multiband observations combining Taiji and 3G ground-based detectors, taking into account the effects of redshift, mass ratio, and inclination angle. Building on this, we construct two population models for IMBHs and simulate their GW observational samples. We then systematically assess the feasibility and potential of using IMBBH mergers as dark sirens to constrain cosmological parameters under different population assumptions and observing durations.
    
The remainder of this paper is organized as follows. In Section~\ref{Method}, we describe the methodology used to simulate IMBH merger events and their joint observation by Taiji and 3G ground-based detectors. Section~\ref{sec:results and discussion} presents the resulting constraints and discusses their cosmological implications. Finally, Section~\ref{Conclusion} summarizes our main findings.

\section{Method}\label{Method}

\subsection{Source population and GW simulation}\label{sec:simulate GW source}

In this study, we adopt the parameterized population model introduced in Ref.~\cite{Fragione:2022ams} to simulate a population of IMBH binaries. The corresponding volumetric merger rate is given by
\begin{equation}
    \mathcal{R}(z, M_1, q) = k \, \mathcal{N}(\mu_z, \sigma_z) \, M_1^{-\alpha} \, q^{-\beta},
\end{equation}
where $k$ is a normalization constant and $\mathcal{N}(\mu_z, \sigma_z)$ denotes a Gaussian distribution in redshift with mean $\mu_z$ and standard deviation $\sigma_z$. To explore different astrophysical scenarios, we consider two representative sets of population parameters. The first set, $\{\mu_z,\sigma_z,\alpha,\beta\}=\{2,1,1,1\}$, corresponds to a merger history dominated by repeated hierarchical mergers in dense stellar environments \cite{Fragione:2022egh}, while the second set, $\{\mu_z,\sigma_z,\alpha,\beta\}=\{5,1,1,1\}$, describes IMBH binaries originating from Population~III remnants, for which the merger rate peaks at higher redshifts \cite{Hijikawa:2022qug}.

For each binary, the remaining source parameters are sampled independently and uniformly within their physically allowed ranges. These parameters include the sky location, inclination, polarization, phase, and coalescence time. In particular, the coalescence time $t_{\rm c}$ is sampled uniformly over the mission duration. This population model forms the basis for generating the mock IMBH merger catalogs used in our GW simulations \cite{Dong:2025ikq}.

In the frequency domain, for these simulated sources, the GW strain measured by the detector is given by
\begin{equation}
\bm{\tilde{h}}(f) = e^{-i\Phi} \bm{\hat{h}}(f),
\end{equation}
with
\begin{equation}
\tilde{h}(f) = h_+(f)\,F_{+}(f) + h_\times(f)\,F_{\times}(f),
\end{equation}
where $h_+(f)$ and $h_\times(f)$ are the two GW polarization components generated using the inspiral--merger--ringdown waveform model IMRPhenomD~\cite{Husa:2015iqa,Khan:2015jqa} with the \texttt{pycbc} package~\cite{DupletsaHarms2023}, and $F_{+}(f)$ and $F_{\times}(f)$ denote the corresponding antenna pattern functions of the detector. We use the space-based detector Taiji and the three 3G ground-based detectors, including the Einstein Telescope (ET) and the Cosmic Explorer in the United States (CE1) and Australia (CE2). The antenna pattern functions for Taiji are taken from Ref.~\cite{Ruan:2020smc}, those for ET from Ref.~\cite{Punturo:2010zz}, and those for CE1 and CE2 from Ref.~\cite{LIGOScientific:2016wof}. 

In this work, the relation between redshift and luminosity distance is determined assuming a flat $\Lambda$CDM cosmology, with a Hubble constant $H_0 = 67.27\ \mathrm{km\,s^{-1}\,Mpc^{-1}}$ and a present-day matter density parameter $\Omega_{\rm m} = 0.3166$, in agreement with the Planck 2018 results~\cite{Planck:2018vyg}.

\begin{figure}[htbp]
    \centering
    \includegraphics[width=0.9\linewidth]{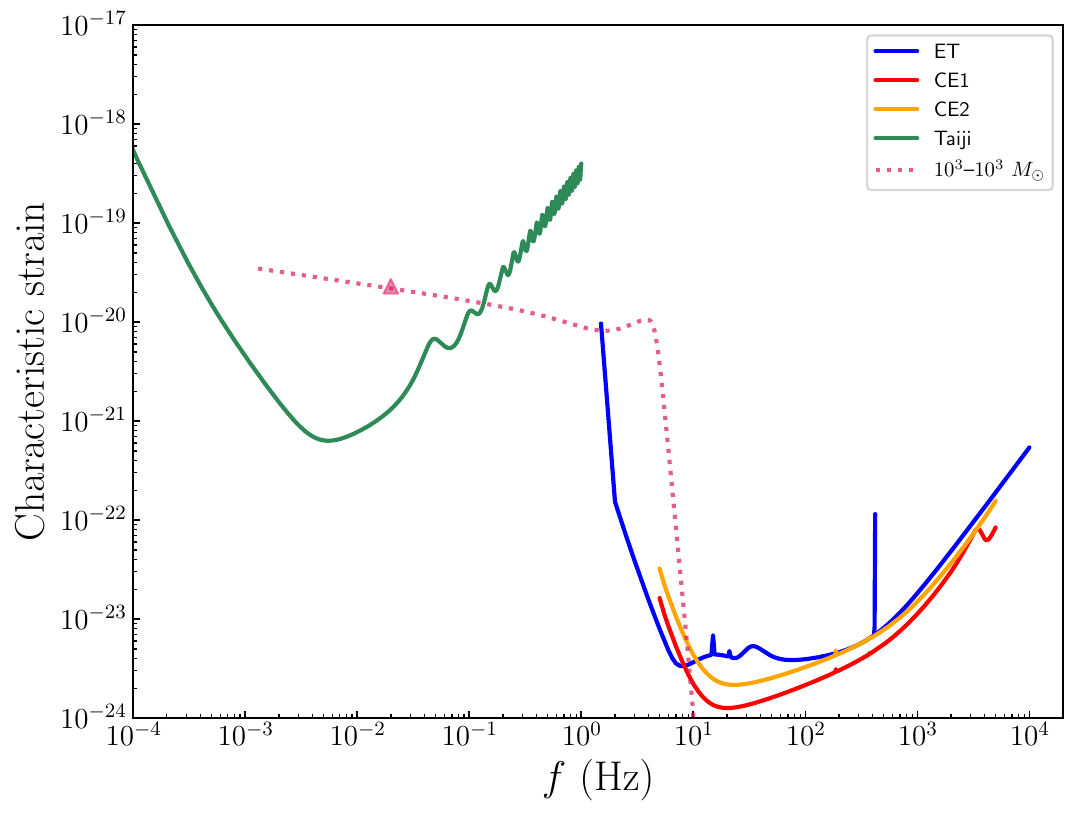}
    \caption{
    Characteristic strains of Taiji, ET, CE1, and CE2, together with that of a merging IMBH binary with component masses of $1000$--$1000\,M_\odot$ at redshift $z=1$.
    The triangle marks the GW frequency one day before coalescence.
    The characteristic strains are defined as $\sqrt{f S_{\rm n}(f)}$ for the detector sensitivity curves and $2f|h(f)|$ for the GW signal.
    }
    \label{fig:fh}
\end{figure}

\subsection{Signal detection and parameter estimation}\label{SNR}

We adopt a signal-to-noise ratio (SNR) detection threshold of 8 for both individual detectors and detector networks, following previous GW studies~\cite{LIGOScientific:2014pky,LIGOScientific:2017adf,LIGOScientific:2016fbo}. For a network of $N$ interferometers, the total SNR is given by
\begin{equation}
    \rho = \sqrt{ \left( \bm{\tilde{h}} | \bm{\tilde{h}} \right) },
\end{equation}
where
\begin{equation}
    \bm{\tilde{h}}(f) = \left[\tilde{h}_1(f), \tilde{h}_2(f), \cdots, \tilde{h}_N(f)\right],
\end{equation}
and the inner product is defined as
\begin{equation}
    \left( \tilde{h} | \tilde{h} \right) = \sum_{k=1}^{N} 4 \int_{f_{\rm in}}^{f_{\rm out}} \frac{\tilde{h}_k(f) \tilde{h}_k^*(f)}{S_{{\rm n},k}(f)} \, df,
\end{equation}
with $S_{{\rm n},k}(f)$ denoting the one-sided power spectral density (PSD) of the $k$th detector and $f_{\rm in}$ and $f_{\rm out}$ corresponding to the frequencies at which the signal enters and exits the sensitive band. We adopt the latest PSD for Taiji, while the PSDs for ET and CE2 are taken from Ref.~\cite{Hild:2010id}, and that for CE1 is taken from Ref.~\cite{CE-web}. The detailed curves are shown in Fig.~\ref{fig:fh}.
  
To estimate the uncertainties in the source parameters, we compute the Fisher information matrix (FIM) for the network,
\begin{equation}
F_{ij} = \sum_{k=1}^{N} 
\left( 
\frac{\partial \tilde{h}_{k}}{\partial \theta_i} 
\Bigg| 
\frac{\partial \tilde{h}_{k}}{\partial \theta_j} 
\right),
\end{equation}
where the parameter vector is
\[
\boldsymbol{\theta} = 
\{ d_{\rm L},\, t_{\rm c},\, \mathcal{M}_{\rm c},\, \eta,\, 
   \theta,\, \phi,\, \psi,\, \iota,\, \psi_{\rm c} \} .
\]
These source parameters are drawn from the population model described in Section~\ref{sec:simulate GW source}.

Inverting the FIM yields the covariance matrix ($\mathrm{Cov}$), from which the $1\sigma$ uncertainties are obtained as
\begin{equation}
\Delta \theta_i = \sqrt{\mathrm{Cov}_{ii}}.
\end{equation}
We then compute the sky localization uncertainty as
\begin{equation}
\Delta\Omega = 2\pi |\sin{\theta}|\sqrt{(\Delta\theta)^2(\Delta\phi)^2 - (\Delta\theta\Delta\phi)^2}.
\end{equation}
The total uncertainty in the luminosity distance combines instrumental, weak-lensing, and peculiar-velocity contributions,
\begin{equation}\label{eq:sigma_dl}
\Delta d_{\rm L} = \sqrt{(\Delta d_{\rm L}^{\rm inst})^2 + (\Delta d_{\rm L}^{\rm lens})^2 + (\Delta d_{\rm L}^{\rm pv})^2} ,
\end{equation}
where $\Delta d_{\rm L}^{\rm inst}$ is estimated from the FIM, the weak-lensing contribution is taken from~\cite{Hirata:2010ba,Tamanini:2016zlh},
\begin{equation}
\Delta d_{\rm L}^{\rm lens}(z) = d_{\rm L} \times 0.066 \left[ \frac{1 - (1+z)^{-0.25}}{0.25} \right]^{1.8} ,
\end{equation}
and the peculiar-velocity contribution is given by~\cite{Kocsis:2005vv}
\begin{equation}
\Delta d_{\rm L}^{\rm pv}(z) = d_{\rm L} \times \left[ 1 + \frac{c(1 + z)^2}{H(z) d_{\rm L}(z)} \right] \frac{\sqrt{\langle v^2 \rangle}}{c} ,
\end{equation}
with the peculiar velocity set to $\sqrt{\langle v^2 \rangle} = 500\ \rm km\,s^{-1}$~\cite{He:2019dhl}.

\subsection{Dark siren cosmological parameter inferences}\label{ds}

The GW dark siren approach infers cosmological parameters by statistically associating each GW event with the redshift distribution of potential host galaxies within its three-dimensional localization volume, using redshift information provided by galaxy survey catalogs. To this end, we construct a mock galaxy catalog by uniformly sampling galaxies in comoving volume with a number density of $0.02~{\rm Mpc}^{-3}$, which corresponds to the median value reported in Ref.~\cite{Barausse:2012fy}, and adopt a conservative apparent magnitude threshold of 24.5, corresponding to the expected depth of future wide-field imaging surveys such as the Legacy Survey of Space and Time \cite{LSST:2008ijt} and the China Space Station Telescope \cite{CSST:2025ssq}. For the redshift uncertainties of galaxies in the mock galaxy catalog, we adopt the optimistic assumption that the galaxy redshift uncertainty is given by $\Delta z(z) = 0.02(1+z)$, following Refs.~\cite{Gong:2019yxt,Song:2022siz,Cao:2017ph}.

For each GW event, the three-dimensional localization region is approximated as a truncated cone in luminosity distance and sky position. The radial range is defined as
\[
[d_{\rm L}^{\rm min}, d_{\rm L}^{\rm max}] = [\bar{d}_{\rm L}-3\Delta d_{\rm L}, \bar{d}_{\rm L}+3\Delta d_{\rm L}],
\]
while the angular region is specified by the condition
\[
\chi^2 = (\theta-\bar{\theta},\phi-\bar{\phi})\, Cov'^{-1}
\begin{pmatrix}
\theta-\bar{\theta} \\
\phi-\bar{\phi}
\end{pmatrix}
\leq 9.21 .
\]
Here, $\bar{d}_{\rm L}$ and $\Delta d_{\rm L}$ denote the mean luminosity distance and its $1\sigma$ uncertainty, respectively, and $Cov'$ is the $2\times2$ covariance matrix of $(\theta,\phi)$. The criterion $\chi^2 \leq 9.21$ corresponds to the $99\%$ confidence region~\cite{Muttoni:2023prw}. To associate GW events with the galaxy catalog, the radial distance interval of each event is further converted into a corresponding redshift range, following the procedure described in Ref.~\cite{Dong:2024bvw}.

We then perform Bayesian inference on the cosmological parameters ${\bm \Omega}$ using the observed GW events. The posterior distribution is given by
\begin{equation}
    p({\bm \Omega}|\{D_{\rm GW}\}) \propto p(\{D_{\rm GW}\}|{\bm \Omega})\, p({\bm \Omega}),
\end{equation}
where $\{D_{\rm GW}\}$ denotes the GW dataset and $p({\bm \Omega})$ represents the prior distribution of the cosmological parameters. We adopt uniform priors for all cosmological parameters, with
$H_0 \in [20,140]~{\rm km\,s^{-1}\,Mpc^{-1}}$ and $\Omega_{\rm m} \in [0.1,0.5]$.
Assuming that individual GW events are statistically independent, the likelihood function factorizes as
\begin{equation}
\begin{aligned}
    p(\{D_{\rm GW}\}|{\bm \Omega})
    &= \prod_{i=1}^{N_{\rm GW}} \frac{1}{\beta({\bm \Omega})}
    \int \int p(D_{{\rm GW},i}|d_{{\rm L},i}) \\
    &\quad \times \delta\!\left(d_{{\rm L},i}-d_{\rm L}(z_i,{\bm \Omega})\right)
    \, p(z_i|{\bm \Omega}) \,
    {\rm d}d_{{\rm L},i}\,{\rm d}z_i ,
\end{aligned}
\end{equation}
where $p(D_{\rm GW}|d_{\rm L})$ is assumed to be a Gaussian distribution centered at $d_{\rm L}$ with a standard deviation $\Delta d_{\rm L}$.

The redshift prior $p(z|{\bm \Omega})$ accounts for both the angular consistency between galaxies and the GW localization, as well as the incompleteness of the galaxy catalog, and is expressed as
\begin{equation}
\begin{aligned}
p(z|{\bm \Omega}) = \bigg\{ &
\frac{1}{N_{\rm in}}\sum_{j=1}^{N_{\rm in}} w_j \,
\mathcal{N}(z \,|\, \hat{z}_j, \Delta z_j) \, p({\rm G}|z, {\bm \Omega}) \\
& +\, p({\rm \bar{G}}|z, {\bm \Omega})
\bigg\}\mathcal{R}(z).
\end{aligned}
\end{equation}
Here, each galaxy is assigned an angular weight $w_j$ that quantifies the consistency of its sky position with the GW localization~\cite{Muttoni:2023prw}, which is calculated as
\begin{equation}
w_j \propto \frac{1}{2\pi \sqrt{|Cov'|}}
\exp\left[-\frac{1}{2}(\theta-\bar{\theta},\phi-\bar{\phi})Cov'^{-1}
\begin{pmatrix}
\theta-\bar{\theta}\\
\phi-\bar{\phi}
\end{pmatrix}\right].
\end{equation}
In addition, $\mathcal{N}(z|\hat{z}_j,\Delta z_j)$ denotes a Gaussian distribution with mean $\hat{z}_j$ and standard deviation $\Delta z_j$, while $\mathcal{R}(z)$ represents the merger rate of IMBBHs.

The terms $p({\rm G}|z,{\bm \Omega})$ and $p({\rm \bar{G}}|z,{\bm \Omega})$ describe the completeness and incompleteness of the galaxy catalog at redshift $z$, respectively. The completeness $p({\rm G}|z,{\bm \Omega})$ is derived from $P({\rm G}|d_{\rm L},{\bm \Omega})$ via the distance–redshift relation, while the incompleteness is given by $p({\rm \bar{G}}|z,{\bm \Omega}) = 1 - p({\rm G}|z,{\bm \Omega})$. 

To estimate $P({\rm G}|d_{\rm L},{\bm\Omega})$, we mock galaxy luminosities assuming a Schechter function \cite{Schechter:1976iz},
\begin{equation}
p(L)\propto \left(L / L^{*}\right)^{\alpha} \exp \left(-L / L^{*}\right) \mathrm{d} L / L^{*},
\end{equation}
with $\alpha=-1.07$ and $L^* = 1.2\times10^{10}h^{-2}L_{\odot}$, adopting a lower cutoff of $0.001L^*$ following Ref.~\cite{Gray:2019ksv}. The luminosities are then converted to apparent magnitudes, from which $P({\rm G}|d_{\rm L},{\bm\Omega})$ is estimated as the fraction of galaxies falling below the survey threshold, assuming for simplicity that the catalog completeness is uniform across the sky.
The GW selection effect $\beta(\bm{\Omega})$ \cite{Chen:2017rfc} is computed as  
\begin{equation}
\beta(\bm{\Omega})=\int p_{\rm det}^{\rm GW}(d_{\rm L}(z,\bm{\Omega}))\,p(z|{\bm \Omega})\,{\rm d}z,
\end{equation}  
where $p_{\rm det}^{\rm GW}(d_{\rm L})$ denotes the probability of detecting a GW event at distance $d_{\rm L}$.

\section{Results and discussion}\label{sec:results and discussion}

In this section, we first investigate the detectability of IMBBH mergers with Taiji alone and with the Taiji--ET2CE multiband configuration, and compare the observational outcomes for different population models. We then examine the potential of IMBBHs as dark sirens for cosmology and analyze how variations in the population model, which affect the merger rate, influence the resulting constraints on cosmological parameters.

\begin{figure}[htbp]
    \centering
    \includegraphics[width=0.9\linewidth]{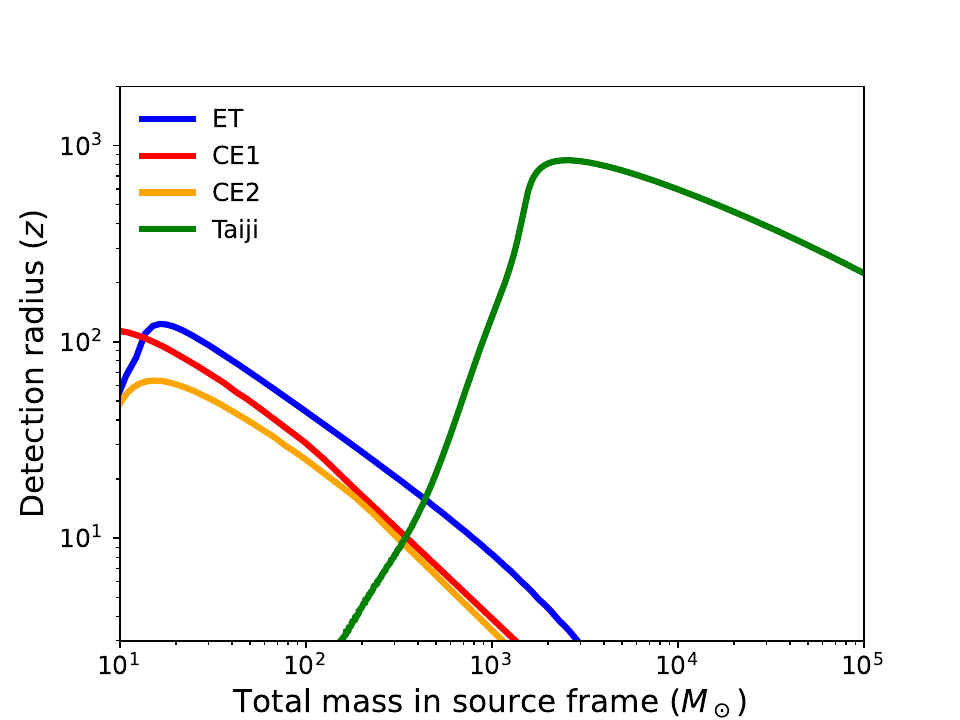}
    \caption{
    Detection horizons of equal-mass, non-spinning binary black holes as a function of the total source-frame mass, for the Taiji, ET, CE1, and CE2 detectors.
    }
    \label{fig:horizen}
\end{figure}

We first assess the detection capabilities of different detectors and discuss the prospects for multiband observations. In Fig.~\ref{fig:horizen}, we show the detection horizons of Taiji and 3G ground-based GW detectors for equal-mass, non-spinning binary black hole systems with an SNR threshold of 8. We see that Taiji broadly covers the mass range of IMBBHs, while 3G ground-based GW detectors are primarily sensitive to lighter binaries, enabling multiband observations. Among the ground-based detectors, ET extends to lower frequencies than CE, thereby reaching higher total masses, as also shown in Fig.~\ref{fig:fh}.

\begin{figure*}[htbp]
    \centering

    \begin{minipage}{0.47\textwidth}
        \includegraphics[width=\linewidth]{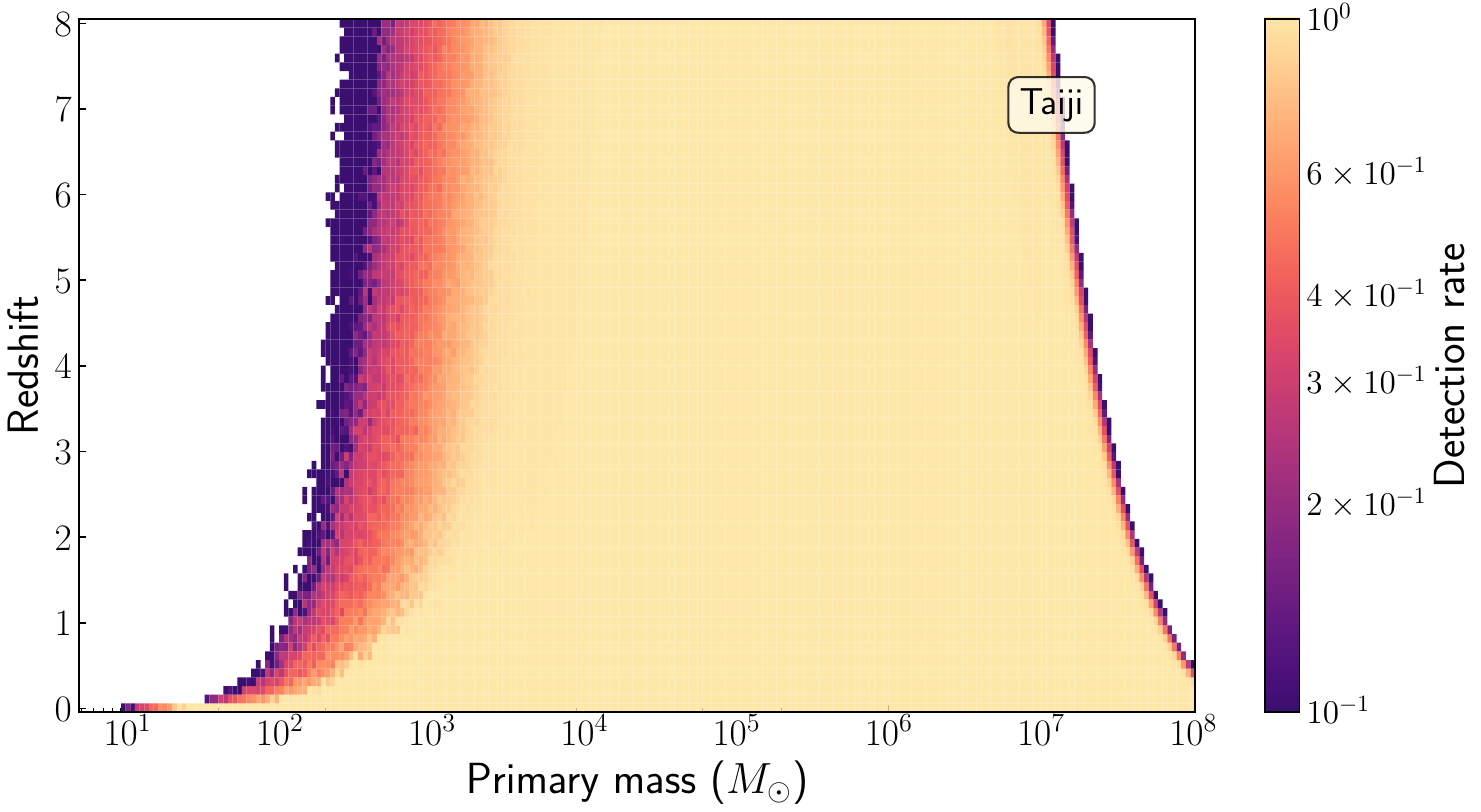}
        \centering
    \end{minipage}
    \hfill
    \begin{minipage}{0.47\textwidth}
        \includegraphics[width=\linewidth]{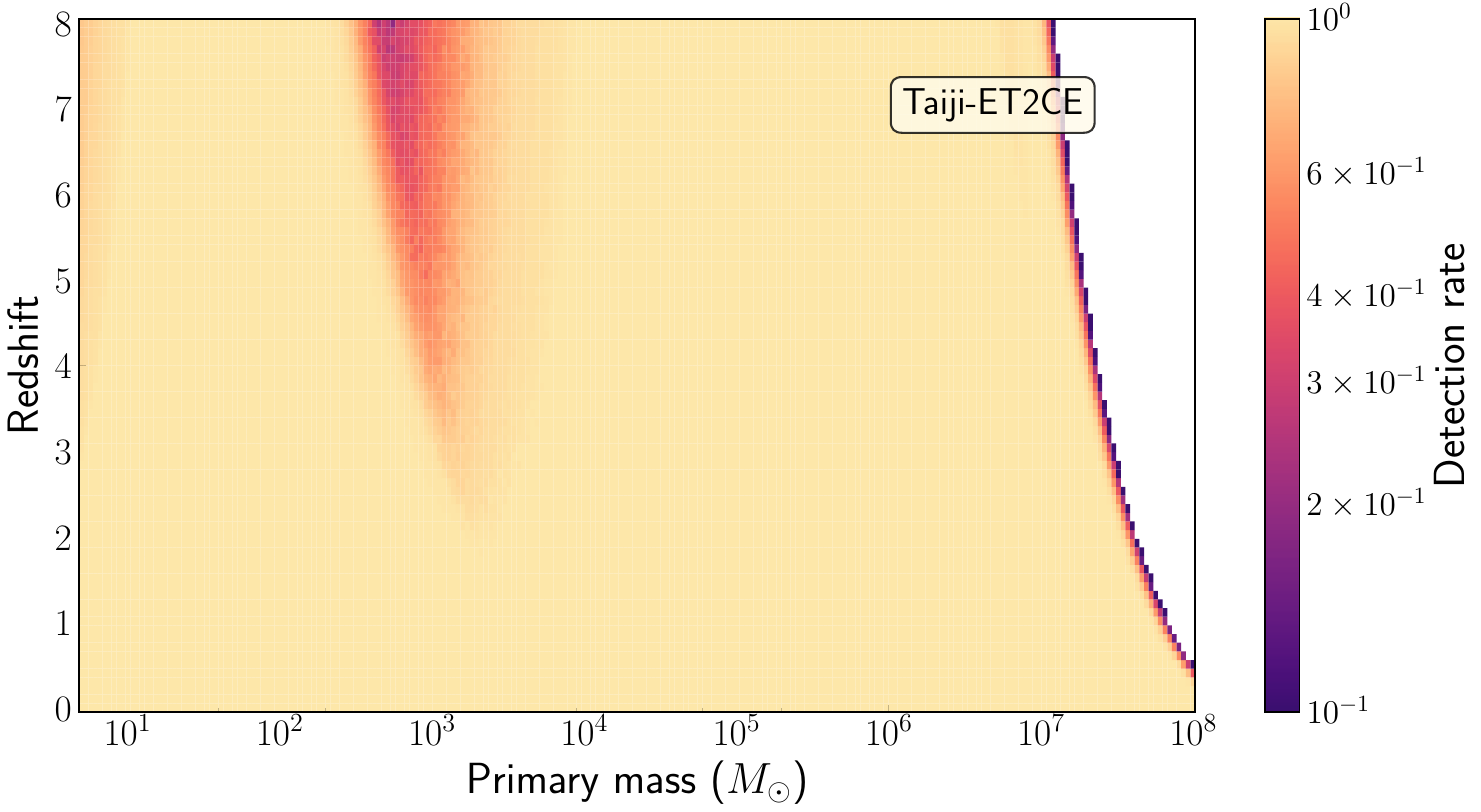}
        \centering
    \end{minipage}

    \vspace{2mm}
    \begin{minipage}{0.47\textwidth}
        \includegraphics[width=\linewidth]{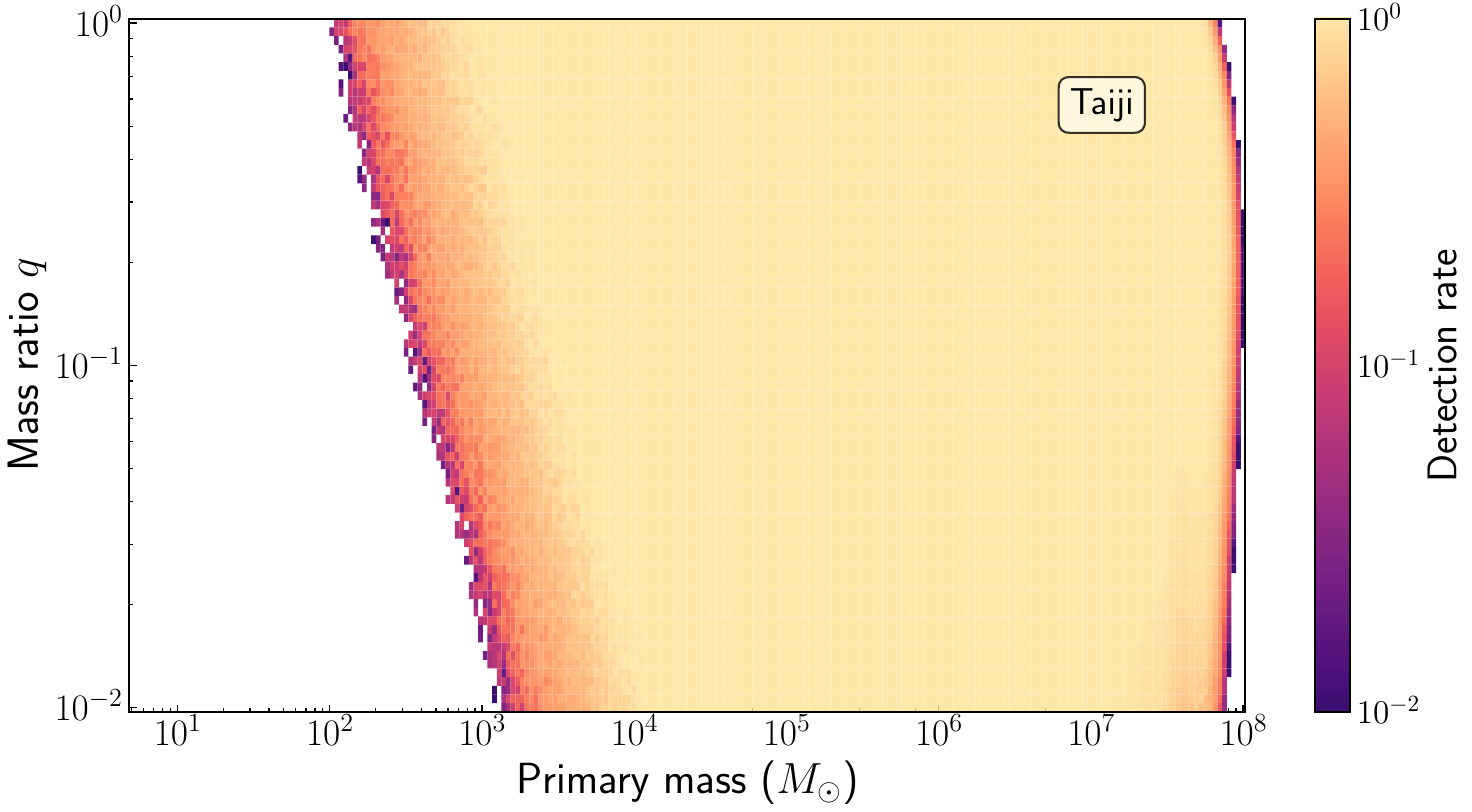}
        \centering
    \end{minipage}
    \hfill
    \begin{minipage}{0.47\textwidth}
        \includegraphics[width=\linewidth]{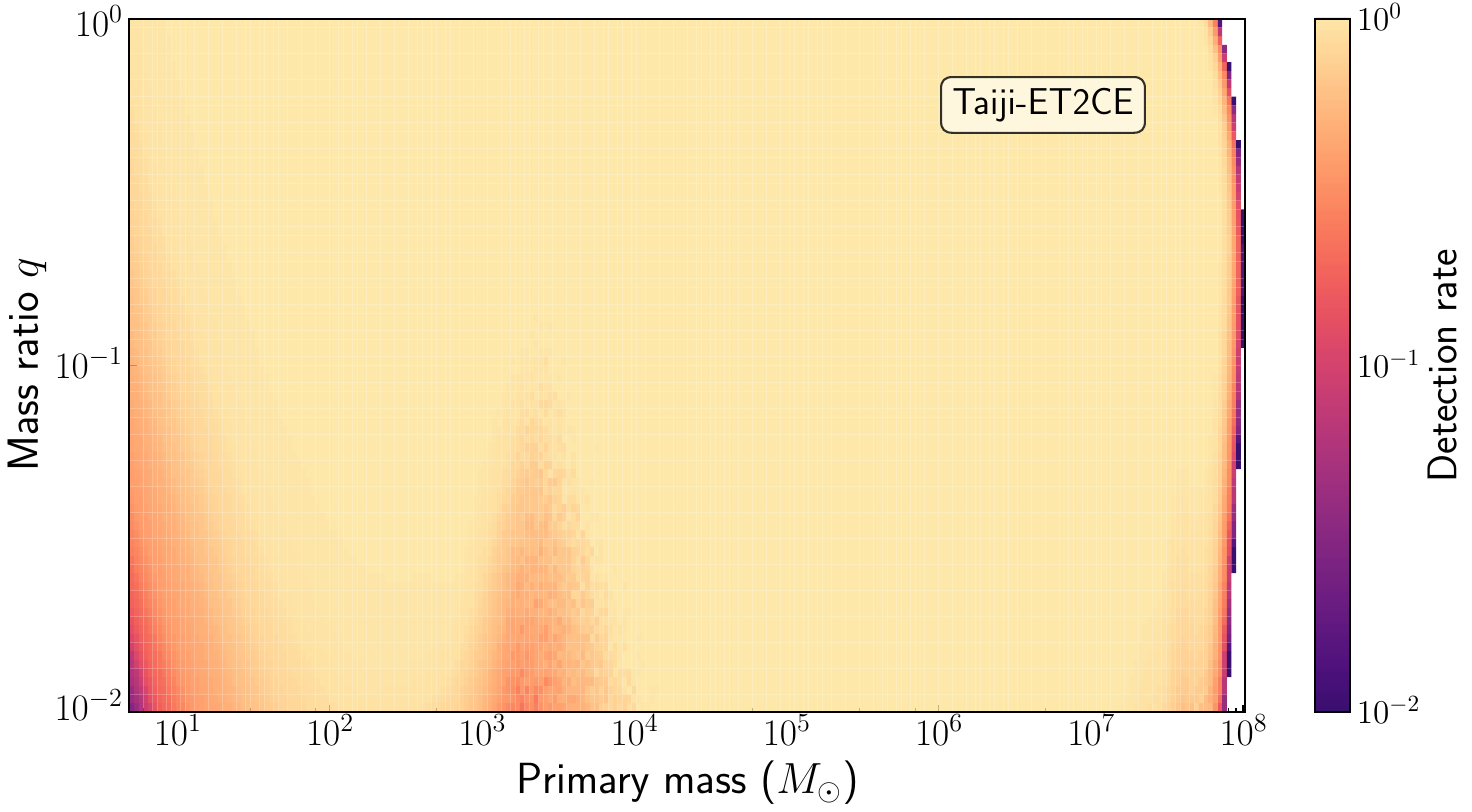}
        \centering
    \end{minipage}

    \caption{Normalized detection capability of black hole binaries in the redshift–primary mass and mass ratio–primary mass parameter spaces for different GW detector configurations. The color scale indicates the relative detection rate.}
    \label{fig:det}
\end{figure*}

To further quantify the detection performance across the parameter space, we present in Fig.~\ref{fig:det} the normalized detection efficiency in the redshift--primary mass and mass ratio--primary mass planes for Taiji and the Taiji--ET2CE multiband configuration.

In the upper panels of Fig.~\ref{fig:det}, we present the detection capabilities of Taiji alone and the Taiji--ET2CE multiband configuration in the redshift--primary mass parameter space for equal-mass binaries with $q=1$. For Taiji alone (left panel), the detector is most sensitive to binary black hole systems with primary masses between $10^3\,M_\odot$ and $10^7\,M_\odot$, maintaining relatively high detection rates up to high redshifts. Outside this mass range, the detection efficiency drops rapidly toward both lower and higher primary masses. When 3G ground-based GW detectors are included (right panel), the Taiji--ET2CE multiband configuration achieves a substantially enhanced detection efficiency in the low-mass regime, with a noticeable reduction occurring only at high redshifts for systems with primary masses close to $10^3\,M_\odot$.
In the lower panels, we show the detection performance in the mass ratio--primary mass parameter space at a fixed redshift of $z=1$. For Taiji alone, sensitivity to the mass ratio is primarily confined to systems with primary masses between $10^2\,M_\odot$ and $10^4\,M_\odot$. As the mass ratio becomes more asymmetric, systems with primary masses between $10^2\,M_\odot$ and $10^3\,M_\odot$ progressively fall below the detection threshold, while the detection efficiency for systems with primary masses between $10^3\,M_\odot$ and $10^4\,M_\odot$ also steadily decreases. By contrast, including ground-based detectors significantly enhances the detectability of low-mass systems with asymmetric mass ratios, with only a limited subset of highly asymmetric binaries with primary masses between $10^3\,M_\odot$ and $10^4\,M_\odot$ exhibiting a partial reduction in detection efficiency.

Overall, these results demonstrate that multiband observations can effectively compensate for the limited sensitivity of Taiji in the low-mass regime, allowing the reliable detection of light IMBHs that would otherwise remain inaccessible, especially those at higher redshifts and with more asymmetric mass ratios, and thereby substantially expanding the parameter space of observable binary black hole systems.

\begin{figure*}[htbp]
    \centering

    \begin{minipage}{0.47\textwidth}
        \includegraphics[width=\linewidth]{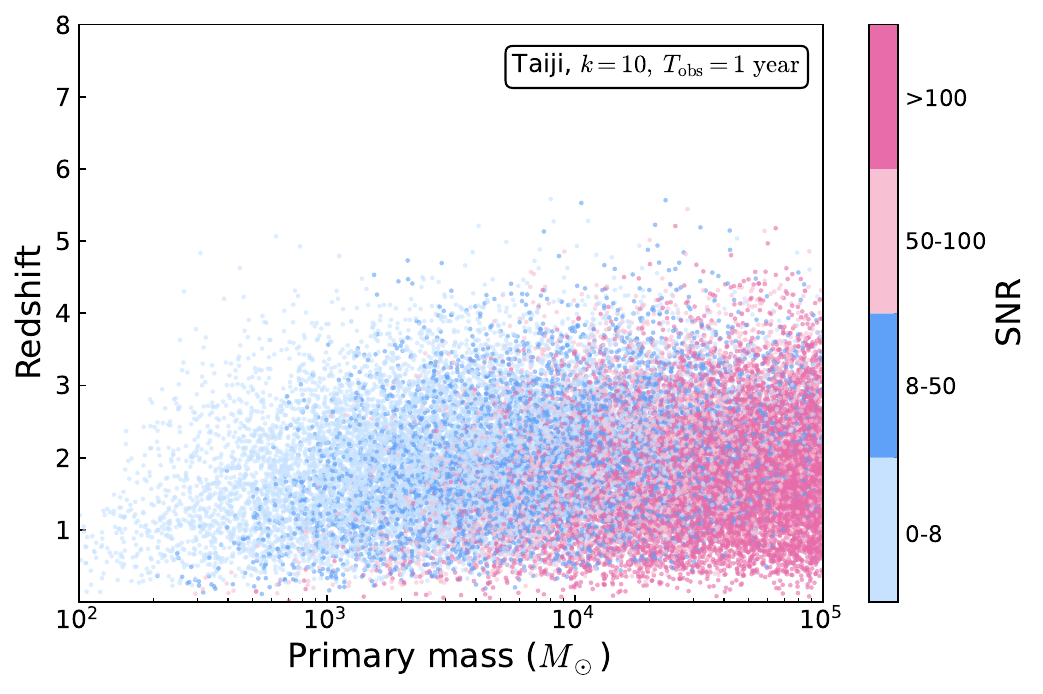}
        \centering
    \end{minipage}
    \hfill
    \begin{minipage}{0.47\textwidth}
        \includegraphics[width=\linewidth]{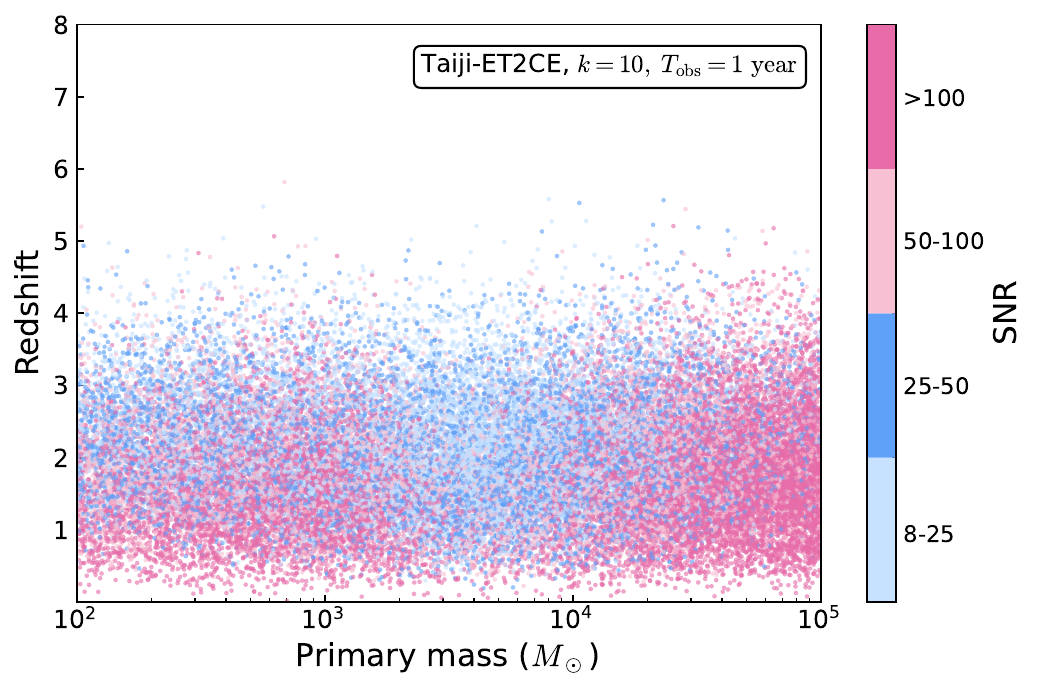}
        \centering
    \end{minipage}

    \caption{Event distributions and corresponding SNRs of IMBBHs for one year of observation with $k = 10$, using $\{\mu_z, \sigma_z, \alpha, \beta\} = \{2, 1, 1, 1\}$. Results are shown for Taiji alone and for the Taiji--ET2CE multiband configuration. The color of each point corresponds to the SNR of the event.}
    \label{fig:2111}
\end{figure*}

\begin{figure*}[htbp]
    \centering

    \begin{minipage}{0.47\textwidth}
        \includegraphics[width=\linewidth]{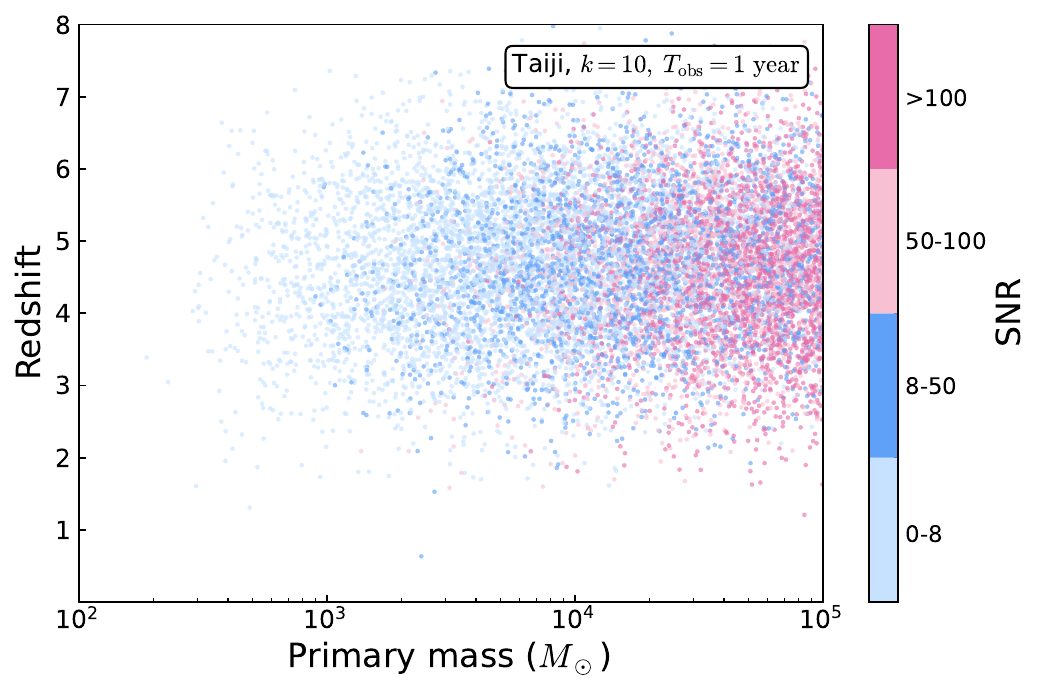}
        \centering
    \end{minipage}
    \hfill
    \begin{minipage}{0.47\textwidth}
        \includegraphics[width=\linewidth]{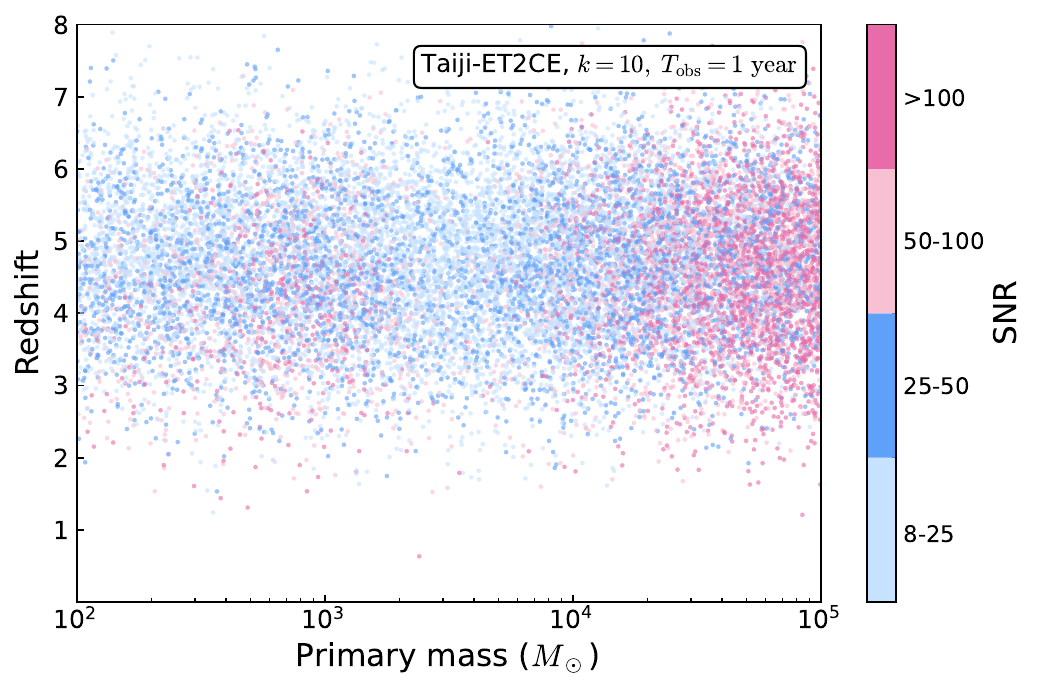}
        \centering
    \end{minipage}

    \caption{Same as Fig.~\ref{fig:2111}, but assuming an intrinsic population model with $\{\mu_z, \sigma_z, \alpha, \beta\} = \{5, 1, 1, 1\}$.}
    \label{fig:5111}
\end{figure*}

Building on the previous detection efficiency analysis, we show in Fig.~\ref{fig:2111}–\ref{fig:5111} the detection prospects of Taiji alone and of the multiband configuration under two different population models, to more clearly illustrate their observational capabilities.
Fig.~\ref{fig:2111} shows the distributions of detectable events and their corresponding SNR over one year of observation. It can be seen that the inclusion of ground-based detectors significantly increases the number of detectable events and enhances their SNR in the low-mass regime, which is broadly consistent with the trends inferred from the detection rates. This allows the multiband configuration to produce high-quality events with SNRs exceeding 100 across almost the entire mass range, which is crucial for subsequent cosmological parameter estimation. As shown in Fig.~\ref{fig:5111}, we observe similar trends, with the total number of detectable events gradually decreasing as redshift increases. In particular, for ground-based detectors, IMBH signals are shifted toward lower frequencies and weaken in amplitude, resulting in a substantial reduction in high-SNR events and thereby limiting their detection capability.

\begin{figure*}[htbp]
    \centering

    \begin{minipage}{0.47\textwidth}
        \includegraphics[width=\linewidth]{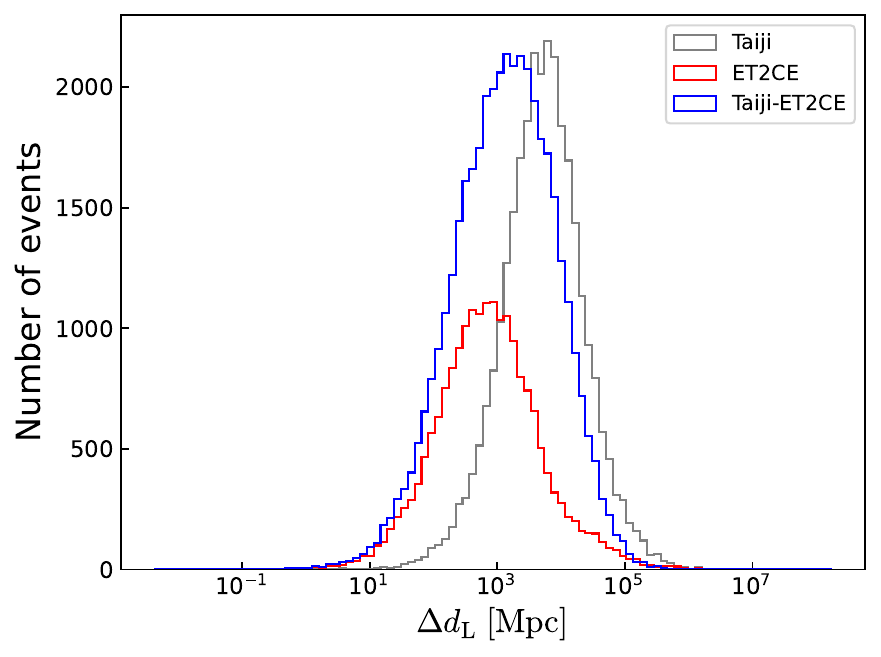}
        \centering
    \end{minipage}
    \hfill
    \begin{minipage}{0.47\textwidth}
        \includegraphics[width=\linewidth]{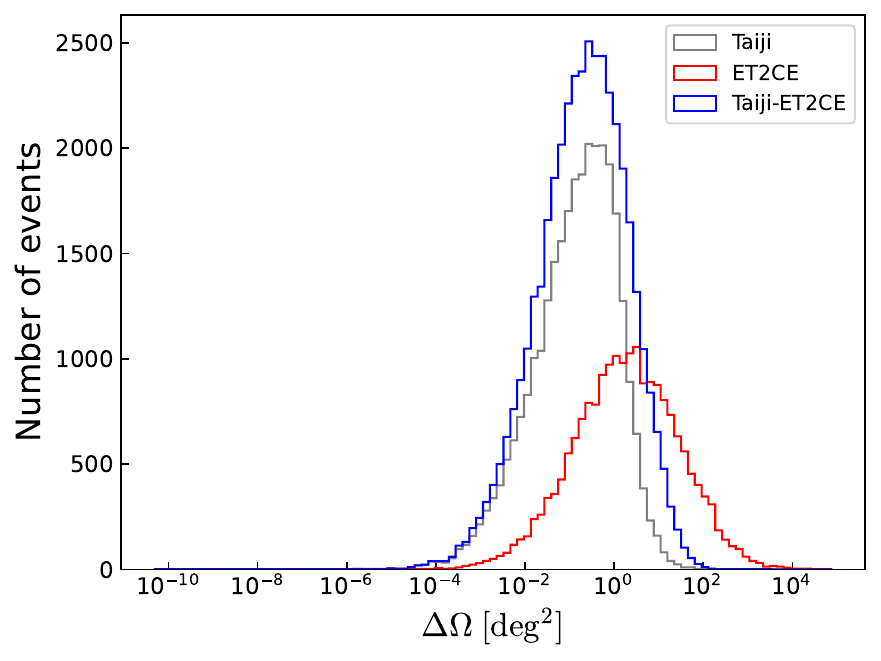}
        \centering
    \end{minipage}

    \caption{Distribution of luminosity distance uncertainty $\Delta d_{\rm L}$ (left) and sky localization area $\Delta \Omega$ (right) for IMBBHs detected by Taiji, ET2CE, and Taiji–ET2CE network for the $\Lambda$CDM model with $k = 10$, adopting $\{\mu_z, \sigma_z, \alpha, \beta\} = \{2, 1, 1, 1\}$.}
    \label{fig:domgea}
\end{figure*}

Having established the detection prospects, we now assess the cosmological potential of IMBBHs as dark sirens. Fig.~\ref{fig:domgea} shows the distributions of uncertainties in luminosity distance and sky localization for the three detector configurations. Taiji achieves better sky localization because it mainly operates in the low-frequency band, observing the inspiral phase of the signals over long durations, with directional modulation induced by its orbital motion enhancing localization. Ground-based detectors provide more precise measurements of the luminosity distance because they are more sensitive to the high-frequency merger signals, which carry richer distance information. Multiband observations with the combined Taiji–ET2CE network significantly enhance localization precision and increase the number of well-localized events, clearly demonstrating the strong complementarity between space-based and ground-based detectors.

\begin{figure*}[htbp]
    \centering
    \includegraphics[width=0.65\linewidth]{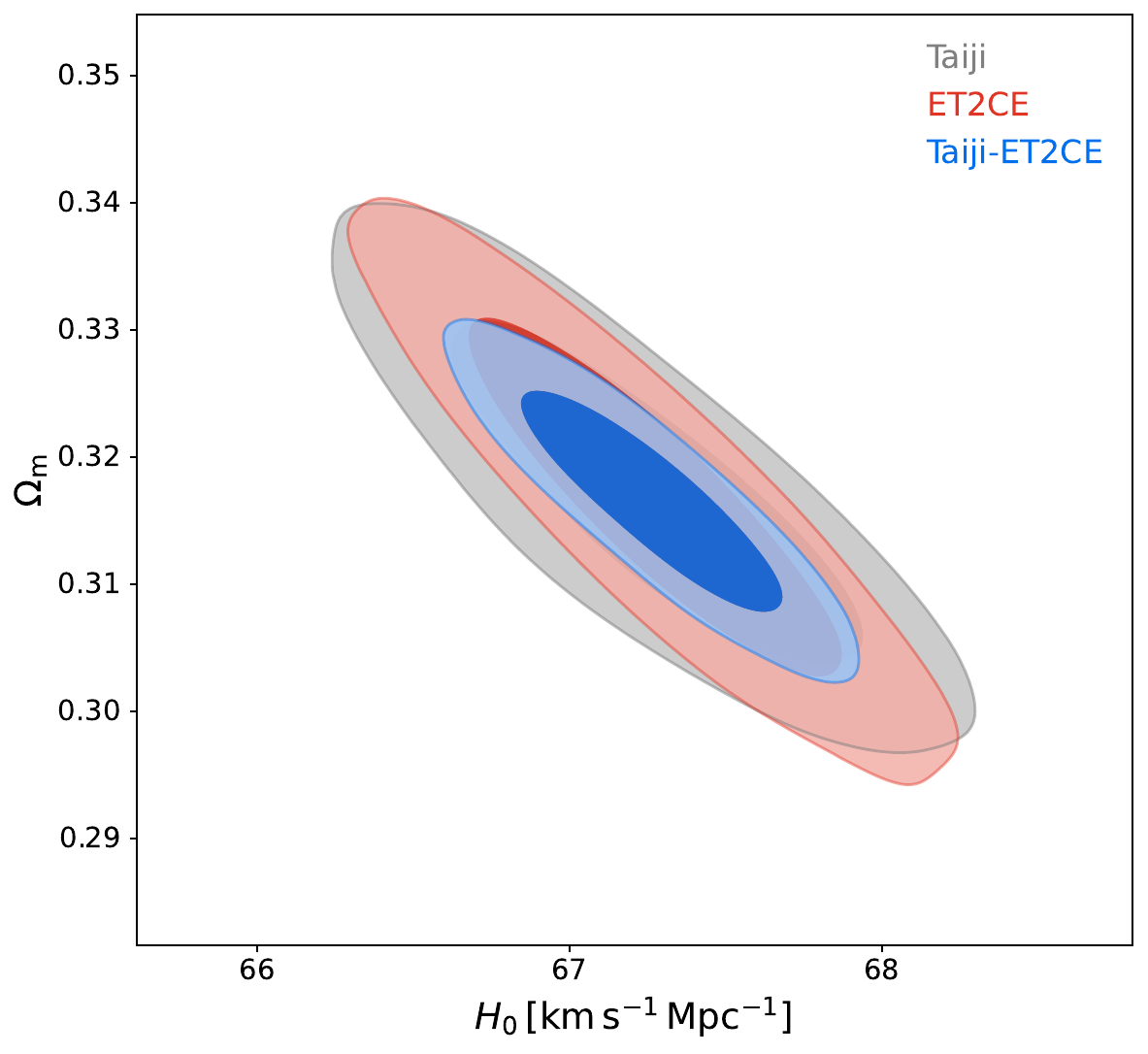}
    \caption{Two-dimensional marginalized contours at 68.3\% and 95.4\% confidence levels in the $\Omega_{\rm m}$--$H_0$ plane from Taiji, ET2CE, and Taiji--ET2CE mock data.
    }
    \label{fig:conta}
\end{figure*}

Based on the localization results, we then carried out an analysis of constraints on cosmological parameters. Fig.~\ref{fig:conta} presents the $H_0$–$\Omega_{\rm m}$ constraints for the three detector configurations shown in Fig.~\ref{fig:2111}, namely Taiji, ET2CE, and the combined Taiji–ET2CE network, representing a space-based detector, a ground-based detector network, and a multiband GW synergetic network, respectively. Using Taiji alone, the constraint precisions on $H_0$ and $\Omega_{\rm m}$ are approximately 0.63\% and 2.8\%, respectively, while the corresponding precisions for ET2CE are 0.58\% and 2.9\%. The multiband Taiji--ET2CE configuration provides significantly tighter constraints, with precisions of 0.40\% and 1.8\%, improving the precision on $H_0$ by 36.5\% and 31.0\% and on $\Omega_{\rm m}$ by 35.7\% and 37.9\% relative to Taiji and ET2CE, clearly illustrating the advantage of multiband GW observations for cosmological parameter estimation. When considering the events shown in Fig.~\ref{fig:5111}, the increase in redshift reduces the number of detectable events and the incompleteness of the galaxy catalog further affects the measurement of cosmological parameters, leading to constraint precisions for the Taiji--ET2CE network of approximately 2.4\% for $H_0$ and 24.4\% for $\Omega_{\rm m}$.

\begin{figure*}[htbp]
    \centering
    \includegraphics[width=0.7\linewidth]{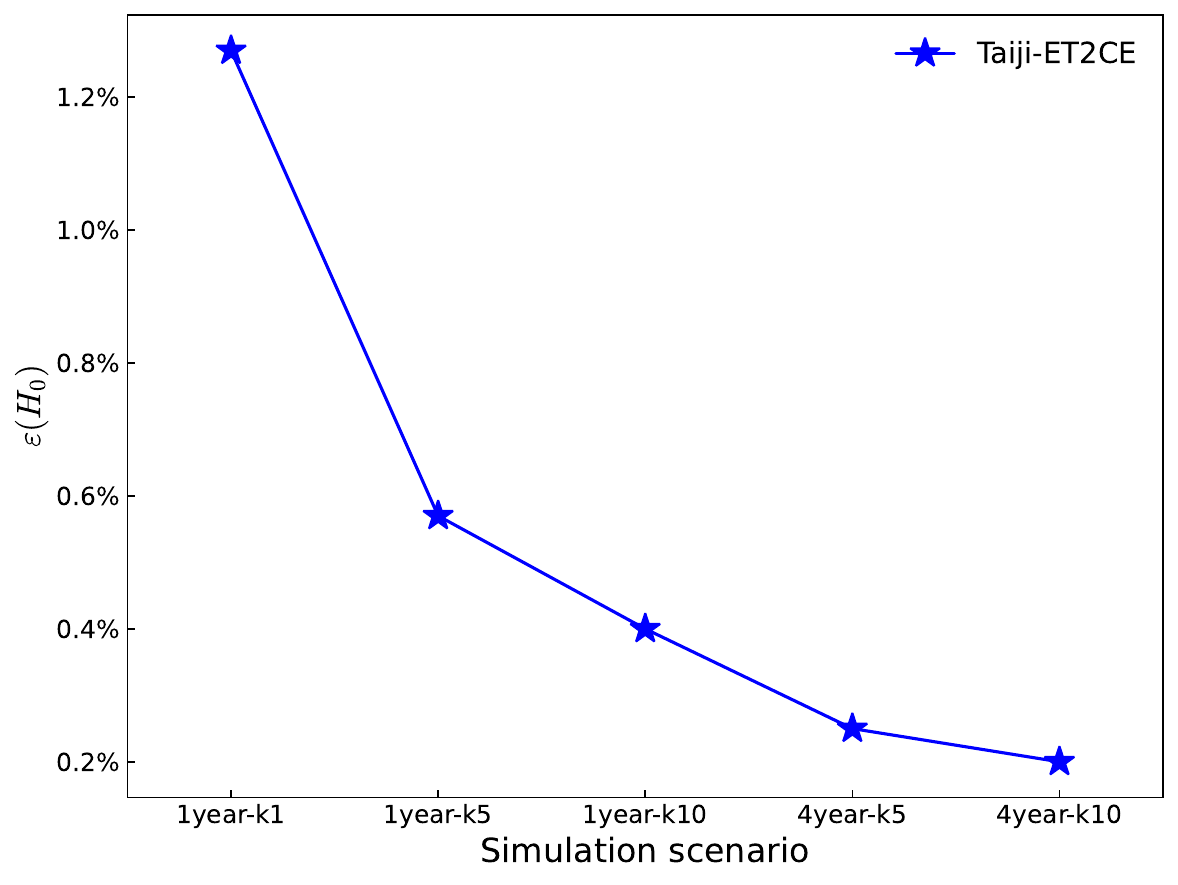}
    \caption{
    Relative errors of $H_{0}$ for different simulation scenarios of IMBBH events simulated with Taiji--ET2CE.
    }
    \label{fig:year}
\end{figure*}

To further illustrate how cosmological constraints depend on the size of the simulated sample, we examine the impact of the number of simulated IMBBH mergers on the precision of the $H_0$ measurement. Fig.~\ref{fig:year} shows the resulting constraints on $H_0$ as a function of the total number of simulated events for the multiband Taiji--ET2CE observations. The results are obtained from different population realizations corresponding to varying observation durations and choices of the normalization factor $k$. As the number of simulated events increases from a few thousand to over one hundred thousand, the precision of $H_0$ measurement improves steadily, tightening from the percent level to the sub-percent level in agreement with the expected statistical gain from the larger number of events. The improvement is particularly pronounced at smaller event numbers, while as the sample size increases further, the gain in parameter constraints gradually becomes less significant.

\section{Conclusion}\label{Conclusion}

The existence of IMBHs, whose masses lie between those of stellar-mass and supermassive black holes, remains largely uncertain in electromagnetic observations, but they are expected to play a key role in black hole formation and galaxy evolution. GW observations provide a direct means to detect IMBH binaries and study their properties, offering the additional possibility of using these systems as dark sirens to probe cosmological parameters.

In this work, we present for the first time a comprehensive analysis of the detection prospects for IMBBHs with the space-based GW detector Taiji and a multiband network including 3G ground-based detectors. We also assess the potential of IMBHs as dark sirens for constraining cosmological parameters. Using two representative population models, we simulate IMBH binary mergers and systematically analyze how their detectability depends on redshift, mass, and mass ratio. Based on the SNRs and sky localization uncertainties derived from the Fisher matrix, we further quantify the effectiveness of these systems in constraining cosmological parameters. In addition, we examine how the number of observed events affects the precision of parameter inference, providing a reference for the potential of future multiband GW networks in astrophysical and cosmological studies.

Our results demonstrate that Taiji alone can efficiently detect high-mass IMBH binaries, whereas adding ground-based detectors significantly enhances the detection of low-mass and asymmetric systems, substantially expanding the observable parameter space. Multiband observations improve SNR and sky localization, enabling precise measurements of source parameters. Consequently, the combined Taiji--ET2CE network significantly improves the precision of both $H_0$ and $\Omega_{\rm m}$ compared with single-detector observations. We also find that increasing the number of simulated events steadily tightens cosmological constraints, with the greatest gains at lower event numbers. 
These results underscore the importance of multiband GW observations of IMBH binaries for both astrophysical studies and precision cosmology, and indicate that future space and ground synergistic networks are expected to provide unprecedented opportunities to probe IMBHs and the expansion history of the Universe.

At the same time, this study has several limitations. We have only considered two relatively simple population models, and our waveform simulations assume nonspinning, circular binaries, neglecting spin, orbital eccentricity, and extreme mass ratios, so not all possible IMBH systems are covered. Incorporating these factors in future work would make the simulations more realistic. 
In addition, this study focuses solely on Taiji as the space-based detector, and future work could explore joint observations with TianQin \cite{TianQin:2015yph}, LISA \cite{LISA:2017pwj}, or even lunar interferometers \cite{LGWA:2020mma, Katsanevas2020LSGA, Amaro-Seoane:2020ahu, Yan:2024jio}, which are expected to further enhance detection capabilities and improve constraints on cosmological parameters, and such joint analyses have already been considered in the context of massive black holes. In particular, TianQin has a sensitivity curve closer to that of ground-based detectors, and joint observations with it may provide improved multiband measurement performance. Overall, these developments are likely to provide stronger support and broader opportunities for future multiband GW networks in both astrophysical and cosmological studies.

\section*{Acknowledgements}
This work was supported by the National Natural Science Foundation of China (Grants Nos. 12575049, 12533001, and
12473001), the National SKA Program of China (Grants Nos. 2022SKA0110200 and 2022SKA0110203), the China Manned Space Program (Grant No. CMS-CSST-2025-A02), and the 111 Project (Grant No. B16009).
\bibliography{gwmultiband}

\end{document}